# Pedestrian Positioning Using WiFi Fingerprints and A Foot-mounted Inertial Sensor

Yang Gu*†, Caifa Zhou†, Andreas Wieser†, Zhimin Zhou*

†Institute of Geodesy and Photogrammetry
ETH Zurich
Zurich, Switzerland
{yang.gu, caifa.zhou, andreas.wieser}@geod.baug.ethz.ch

*College of Electronic Science and Engineering
National University of Defense Technology
Changsha, Hunan, China
zhouzhimin@nudt.edu.cn

*Abstract*—Foot-mounted inertial positioning (FMIP) and fingerprinting based WiFi indoor positioning (FWIP) are two promising solutions for indoor positioning. However, FMIP suffers from accumulative positioning errors in the long term while FWIP involves a very labor-intensive offline training phase. A new approach combining the two solutions is proposed in this paper, which can limit the error growth in FMIP and is free of any offline site survey phase. This approach is realized in the framework of a particle filter, where each particle denotes a potential trajectory of the user and is weighted according to its consistency in signal strength space. Compared with the traditional Gaussian process based approaches, the proposed one has less computational cost and is free from any prior information in the position domain, such as the positions of access points, received signal strengths at certain positions and so on. An experiment is carried out to demonstrate the performance of the proposed approach compared to the traditional Gaussian process based approach.

## I. INTRODUCTION

After years of gaining attention and despite fast development, indoor positioning is still a great challenge. Different from outdoor positioning, where Global Navigation Satellite Systems (GNSS) fulfill the positioning needs of a broad variety of applications, there is no single indoor positioning solution satisfying the requirements of most users. Combining different indoor positioning solutions and techniques seems necessary to obtain the required accuracy, reliability and coverage. In this paper, we focus on the combination of two complementary and very promising indoor positioning solutions: dead-reckoning based on a foot-mounted inertial sensor, and absolute positioning using WiFi fingerprinting.

The core of foot-mounted inertial positioning (FMIP) is an inertial measurement unit (IMU) mounted on the foot of the user and providing data for calculating the position changes of the user [1][2]. This solution is self-contained and has several attractive features: (1) it does not require any pre-installed special infrastructure; (2) it does not require a database of previously determined reference data; (3) it has no coverage limitations in theory. Overall, this solution is easy to deploy even in an unknown environment. Zero-velocity updates (ZUPTs), i.e., detecting and properly introducing short stationary periods of the foot in the positioning algorithm (e.g., in an extended Kalman filter, EKF) can substantially decrease the positioning error growth from cubic in time to linear in time [3]. Nevertheless, the errors of FMIP are still accumulative and unbounded in the long term. Even if the initial position and orientation are perfectly known, the errors may exceed the admissible levels in many applications.

Fingerprinting-based WiFi indoor positioning (FWIP) is attracting attention because WiFi signals and WiFi enabled mobile devices are becoming more and more ubiquitously available. As a result, no additional infrastructure or hardware is needed for FWIP. Under the assumption that similar positions (positions with low Euclidian distances from each other) correspond to similar received signal strengths (RSS) of the WiFi access points (APs), the RSS values observed at an unknown position can be compared to the ones stored in a database for known locations (fingerprints). In a simple realization the unknown position is estimated using the k-nearest neighbor (kNN) algorithm [4]. The main hurdle does not lie in the position estimation (estimation phase), but in the training phase in which the fingerprints associated with known locations are collected and stored in the fingerprinting radio map prior to positioning. Usually the training phase is slow and labor intensive [5]. The efficiency for building the radio map can be improved by crowd sourcing [6][7]. However, the approaches available so far require users to share and upload sensor data from their mobile devices during this phase and are thus still labor intensive.

An existing approach combining the two indoor positioning solutions adopts Gaussian processes (GPs) to model the RSS measurements [8][9]. However, the GP based approach has three problematic features: high computational complexity, inconsistent weight updates, and difficulty in initialization. We therefore propose a new approach inspired by simultaneous localization and mapping (SLAM) [10][11] in this paper. It allows exploiting the RSS measurements even if no radio map exists (yet). The approach is based on a particle filter, where each particle represents a potential trajectory of the user. The particles are weighted according to their consistency in the RSS space, i.e. by assessing the similarity of the respective IMU-based and fingerprinting-based position. As opposed to the existing approach based on GP we predict the position using RSS values rather than predicting the RSS values using positions. We consider this more appropriate because the errors of the RSS measurements are virtually independent of time while the ones of the IMU-based positions grow quickly with time. Indeed we will show that the proposed approach achieves better performance than the GP-based approach, and has further advantages.

The structure of this paper is as follows. The particle filter used in our approach is different from a standard particle filter. Its fundamentals are introduced in Section II. In Section III, the GP-based approach and its disadvantages are discussed, and the proposed weight update strategy is described. The application is demonstrated in Section IV analyzing data from an indoor/outdoor experiment.

## II. Fundamentals of Particle Filter

Particle filters are widely used for non-Gaussian and nonlinear filtering [12]. Fundamentals of particle filtering can be found in e.g., [13] and [14]. In our application we assume that there is no radio map. So, a single RSS observation cannot be used to update weights of particles representing positions, because the likelihood of the RSS observation cannot be calculated. Therefore, some adaptations are made to the particle filter. They are inspired by the SLAM technique, in particular by loop-closure [10].

### A. Structure of particles

In many particle filter based positioning approaches, see e.g. [15], each particle represents a candidate of the state vector representing the position and heading of the user at a particular time. We will only consider two-dimensional positions herein and subsequently use $pos = \{x, y\}$ to denote a position, and $\theta$ to denote the heading. As the idea of loop-closure is applied in our implementation, the history of the state vector is needed as well. Therefore, the particles are extended to represent not only the latest position but also the history of positions. Each particle thus denotes a potential trajectory of the user. Fig. 1 shows the chosen structure of the particles assuming that the history taken into account extends $k$ epochs back in time from the current one, and that the ensemble comprises $N$ particles. For all expressions relating to particles herein, e.g., $pos_j^i$, the superscript denotes the particle index and the subscript denotes the time index.

At each time represented within the particles, the user is at a certain, possibly different, location, and the RSS values observed at that time are stored in a list (also indicated in Fig. 1). As these observations are given, they are not part of the particles but kept in a separate list.

### B. Particle propagation

In our implementation, we assume that FMIP including ZUPTs is available as an encapsulated solution such that the foot-mounted sensor or the computer processing its data outputs a new estimate of position or position increment each time a step is detected by the sensor. Since such algorithms have been published previously, e.g. [2], we will not further elaborate on FMIP here. Instead we assume that the particles are updated with each detected step, and that this update is based on the position increments derived from the FMIP output parameterized as estimated step length $\Delta L$ and heading $\Delta \theta$.

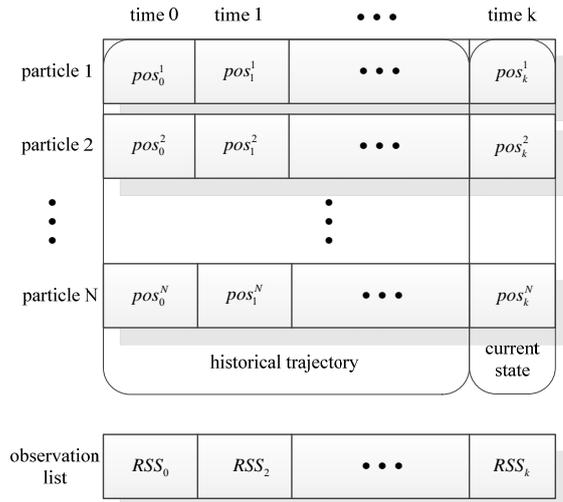

Fig. 1 The structure of the particles, and the observation list. Each particle represents a potential trajectory of the user at k+1 times.

The process of particle propagation is visualized in Fig. 2 which shows three instances with three particles, assuming the initial position and heading are known beforehand (blue triangle and arrow). The $i^{th}$ particle is propagated from time 0 to time 1 by appending a candidate position $pos_1^i$ to the initial position $pos_0$:

$$pos_1^i = pos_0 + (\Delta L_1 + \delta L_1^i) \cdot \begin{pmatrix} \cos \theta_1^i \\ \sin \theta_1^i \end{pmatrix} \quad (1)$$

where

$$\theta_1^i = \theta_0 + \Delta \theta_1 + \delta \theta_1^i \quad (2)$$

$\Delta L_1$ and $\Delta \theta_1$ are the measurements output by the FMIP system (or directly derived therefrom). $\delta L_1^i$ and $\delta \theta_1^i$ are obtained using a random number generator and represent possible corrections for the FMIP measurement noise. These are the contributions creating different particles starting from the same initial position, orientation and measurements, see Fig. 2a. The candidate noise corrections are drawn from the (assumed or known) probability distributions of the FMIP system's step length noise and heading change noise, i.e.,

$$\delta L_1^i \sim p(\varepsilon_{\Delta L}) \quad (3)$$

$$\delta \theta_1^i \sim p(\varepsilon_{\Delta \theta}) \quad (4)$$

Similarly, the propagation from epoch $k-1$ to $k$ consists in appending a candidate position $pos_k^i$ to the positions $pos_0, ..., pos_{k-1}^i$, where

$$pos_k^i = pos_{k-1}^i + (\Delta L_k + \delta L_k^i) \cdot \begin{pmatrix} \cos\theta_k^i \\ \sin\theta_k^i \end{pmatrix} \quad (5)$$

and

$$\theta_k^i = \theta_{k-1}^i + \Delta\theta_k + \delta\theta_k^i \quad (6)$$

This means that we have an ensemble of $N$ particles, each of which represents a potential trajectory of the user from time 0 to time $k$, where each of these particles can then be written as

$$P_k^i = [pos_0, pos_1^i, pos_2^i, ..., pos_k^i] \quad (7)$$

in agreement with Fig. 1. Three such particles are visualized in Fig. 2 for epochs 1, 2 and $k$.

### C. Observation approximation

We have chosen to update the particles and thus obtain a new position estimate each time the FMIP sensor detects a step (*step-wise update*). Since the FMIP sensor may be a black-box sensor operated independently from the WiFi signal strength sensors, and since step duration may vary while RSS measurements may take place at regular time intervals, the RSS observations will usually not be synchronized with particle update, see Fig. 3.

There are several possibilities how to synchronize the measurements computationally. In order to allow real-time processing without waiting for future measurements, extrapolation of the RSS values based on regression or prediction based on time series analysis would be viable options. For simplicity, we just chose the most recent RSS reading before the step here, and assume that the time interval and position change between this RSS measurement and the step is negligible. If, for a particular access point or for all of them, no RSS observation is obtained between the previous step and the current one, the corresponding value(s) are indicated as not available in the list of RSS observations.

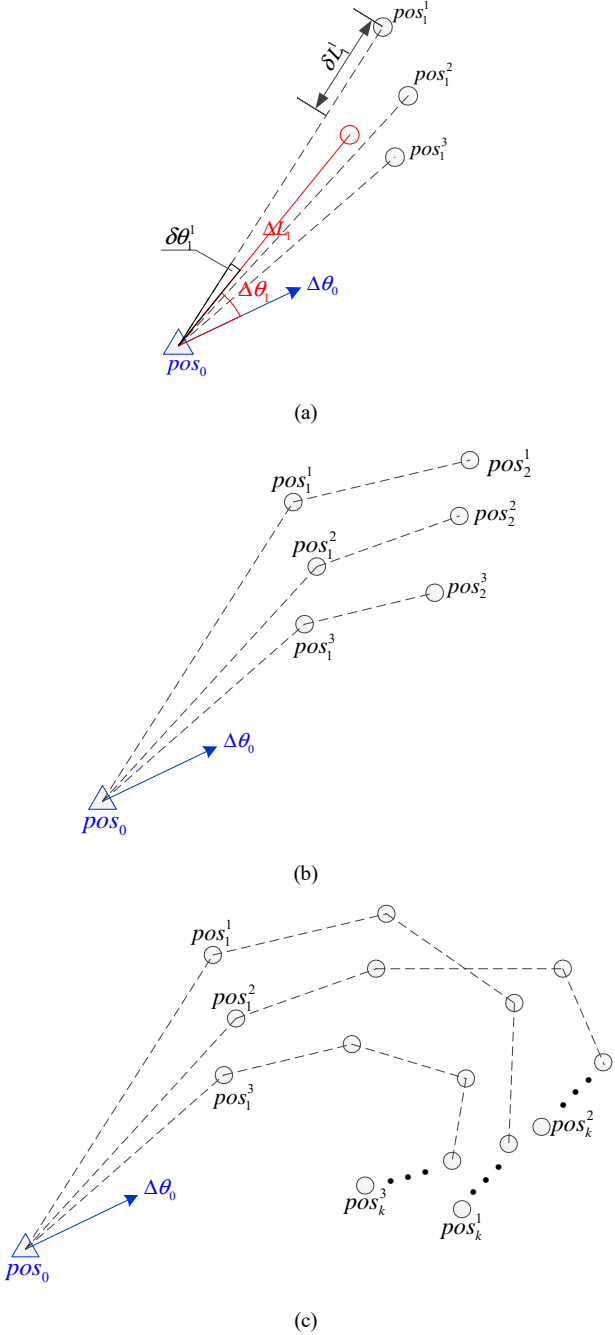

Fig. 2 Three potential trajectories (three particles) at time epoch t=1 (a), t=2 (b), and t=k (c).

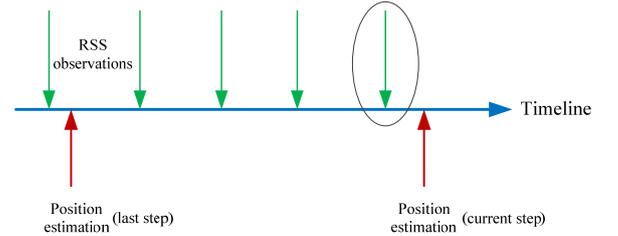

Fig. 3 Alignment of the observed RSS and the position data. In our implementation RSS measurements come from the sensors of a WiFi enabled smart phone and the time is not synchronized with the steps. The step-wise position data comes from the foot module. The RSS observation made closest to the time of the current step is considered as current observation (in the ellipse).

### III. WEIGHT UPDATE STRATEGIES

Within a particle filter, each particle is associated with a weight. Particles with a high weight are more likely to represent the true values than particles with a low weight. In a standard particle filter, the weights are calculated from the likelihood of the observations actually available.

In the GP based approaches, GP regression is carried out to predict the distribution of the RSS in the current location by taking the RSS observations of nearby positions as training samples. Then the weight of each particle can be calculated from the likelihood of the RSS observations derived from the distribution predicted using the coordinates within the particle.

## A. Traditional GP based weight update strategy

A GP can estimate distributions over functions based on training data [16]. It is suitable for modeling signal strength measurements. The advantages are as follows:

- With several training samples, a GP can predict the RSS at arbitrary positions.

- It provides a predictive distribution which is suitable for assessing the quality of the predicted RSS values and for calculating the likelihood of the RSS measurements.

- A GP can cope well with nonlinearity, which is beneficial for RSS-based indoor positioning because the relation between RSS and positions is highly nonlinear due to the complexity of indoor environment.

It is usually assumed that the RSS of different APs are independent and thus GPs for the RSS of each AP can be estimated separately. Details for GP regression have already been described by other authors such as [16] and [17], so they are skipped here. We only use a numeric example of GP regression (see Fig. 4) to highlight some key aspects.

In this example, raw signal strength measurements of one AP are acquired at individual locations (Fig. 4a) by a user walking around a certain area. The positions and RSS values are provided by the foot-mounted IMU module and by the mobile device described in sec. IV, respectively. The synchronization is carried out like depicted in Fig. 3. As the total walking time is short in this case, the position errors are considered insignificant. With a GP, the Gaussian predictive distribution of signal strength at arbitrary positions in the area can be derived. The predicted mean and variance are shown in Figs. 4(b) and 4(c), respectively. Not surprisingly, the areas with few samples, primarily near the borders of the covered area, have larger variance. When applying this concept to particle filters and positioning, the predictive distribution is not needed at each position in the area but only at the current particle's position.

## B. Limitations of the GP based approaches

Although GPs are suitable for modeling signal strength measurements, the existing GP based approaches have three limitations.

- The computational cost is too large. At each filtering epoch, the number of GPs to be trained is the number of particles multiplied by the number of APs (as above, the RSS of different APs are assumed independent). The number of training samples for each of these GPs comprises all nearby positions and the related RSS values. It increases with time. If the user walks around in a small area, the newly collected samples are all "nearby" training samples and the number of training samples grows even faster.

- GPs are by default zero mean processes. So they can only be used here with a proper estimation of the mean values of the RSS, which are non-zero. There are mainly three mean offset models: (1) constant mean (can lead to incorrect predictions at areas with few

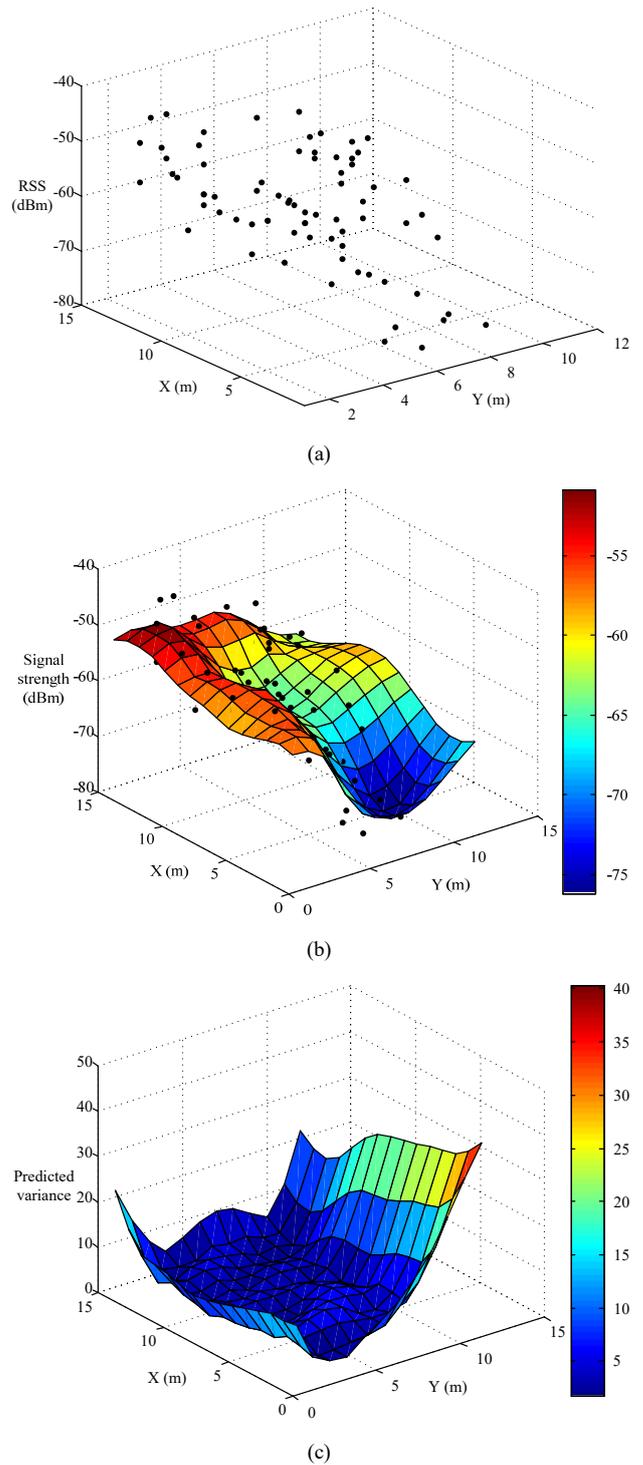

Fig. 4 (a) Raw signal strength measurements for one AP; (b) predicted mean and (c) variance for the entire area obtained from the trained GP.

training samples); (2) model of signal strength decreasing linearly with distance from the AP [9]:

$$mean = k \| p - p_{AP} \| + d \qquad (8)$$

where $k$ is propagation slope, $d$ is the signal strength at the AP, $p$ is the current position and $p_{AP}$ is the AP's position; (3) the log-distance model [8]:

$$mean = s - q \| p - p_{AP} \| \quad (9)$$

where $s$ is the signal strength measured at 1 m from the AP, $q$ is the attenuation factor, $p$ and $p_{AP}$ are as above. Both the second and the third model need additional information such as the positions of all APs and some signal strength measurements at certain positions with known distances from the APs. Implementing this is very labor-intensive, especially when the number of APs is large, and the models may not be sufficient approximations to the true situation in complex indoor environment.

- The weight update of the particles is inconsistent. Fig. 5, shows a constructed example with three particles, each denoting a possible trajectory. The trajectories differ because of the uncertainties due to inertial drifts. The markers denote positions (according to the particles) at which RSS measurements are available. All positions with RSS measurements within a certain predefined radius (see circles in Fig. 5) are considered training samples of the GP for the RSS at the center of the circle. All other positions are excluded because they are too far away from the current position. The reason for excluding samples outside a certain radius is to lower computational cost for online GP applications, see [18], which is also known as sparse approximation of full GP. In this particular example, the number of training samples for the three particles are 1, 2 and 0, respectively. Therefore, the likelihood calculated from the GPs is not consistent for the three particles. This can be partially mitigated by keeping particles with low likelihood [19], however, this introduces the problem of falsely keeping wrong particles.

### C. RSS distance based weight update

As mentioned above, we aim at an algorithm that does not require a radio map and thus we cannot calculate the likelihood of the RSS observations because we do not have expected RSS values to which the actual measurements could be compared. However, with the general assumption that similar RSS values correspond to similar positions, we can calculate weights of the particles by comparing the current RSS values to those obtained at earlier times and stored in the observation list. This is similar to the idea of loop-closure, where correlations between observations can be used to calibrate the estimated positions. This idea also yields an algorithm which is computationally less demanding than the GP-based approach. Finally, the new approach only requires searching nearby RSS values in the RSS space, thus avoiding the search for nearby positions in the coordinate space affected by errors of position estimation which grow rapidly with time.

In the new approach, the weights of the particles are updated according to the consistency in the RSS space. A similarity metric is required for finding similar RSS observations and subsequently assessing consistency. We accomplish this using a normalized Euclidean distance for two RSS vectors $RSS_A$ and $RSS_B$, where:

$$RSS_A = [RSS_A^{(1)}, RSS_A^{(2)}, RSS_A^{(3)}, ..., RSS_A^{(N)}] \\ RSS_B = [RSS_B^{(1)}, RSS_B^{(2)}, RSS_B^{(3)}, ..., RSS_B^{(N)}] \quad (10)$$

The superscripts in the vectors denote the AP to which the respective signal strength refers. Some special precautions are needed if $RSS_A$ and $RSS_B$ contain RSS values from different APs, for example, if the signal strength of a certain AP is available in $RSS_A$ but not in $RSS_B$. Within this paper we have chosen to set such missing RSS values to -110 dBm assuming that the respective signal is just buried within noise and the actual signal strength is very low. (This is a workaround and will further be investigated in the future.)

The normalized Euclidean distance between two RSS vectors of dimension $N$ is then defined as follows:

$$dis(RSS_A, RSS_B) = \sqrt{\frac{\sum_{j=1}^{N}(RSS_A^{(j)} - RSS_B^{(j)})^2}{N}} \quad (11)$$

Fig. 6 shows the distance comparisons in signal space and coordinate space for an experiment where a certain closed path was walked three times. The horizontal axis denotes the epoch index of the collected RSS samples, and the vertical axes denote the normalized Euclidean distance in RSS space (left axis, blue dashed line) and the Euclidean distance in coordinate space (right axis, red line). All the distances are distances from the fiftieth sample. The markers (samples 153 and 265) denote observations made at almost the same place as the fiftieth sample. Both, coordinates and RSS values do not reach exactly the same values again when this place is revisited. However, the drift in position is much larger, and the local minimum distance does not correspond to the actual revisiting of the site – showing the impact of accumulative error growth of the IMU-based positions output by the FMIP system.

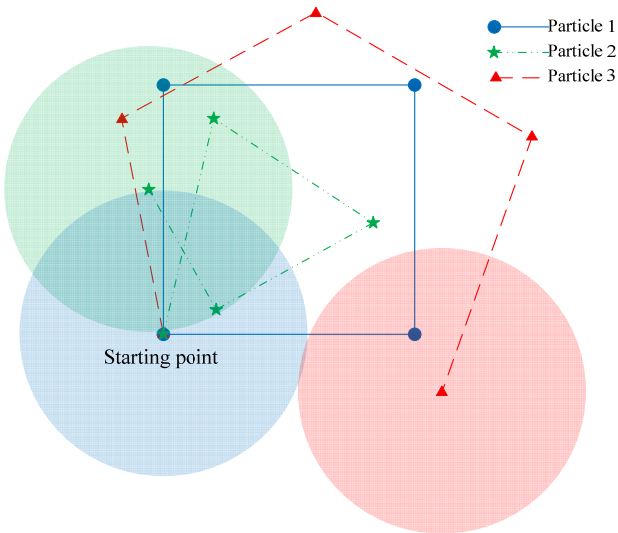

Fig. 5 Simple example to show that different particles have different numbers of training samples and thus likelihoods calculated therefrom are not directly comparable. There are three particles, each particle denotes a possible trajectory. The markers denote positions are which RSS measurements are available. If a position with RSS measurement is within a circle, it is considered one of the training samples of the GP, otherwise it is excluded from GP training.

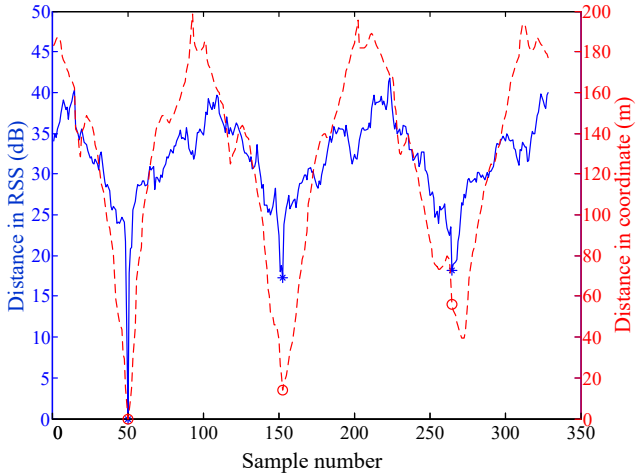

Fig. 6 Distances w.r.t. 50$^{th}$ sample in RSS space (blue) and coordinate space (red) during a repeated closed-loop walk; markers denoting observations made at approximately the same place as the 50$^{th}$ sample.

After finding similar RSS values in the RSS space, we can use the spatial correlations of the RSS to check the estimated positions. Fig. 7 shows the relation between the distances in the two spaces for three different real world data sets: one was acquired outdoors around a building (Fig 7, top), one in the corridors of a building (Fig. 7, center) and the third one (Fig. 7, bottom) in different rooms within in the building. Each data set was acquired during a walk with total walking time less than 5 minutes, such that the position estimation errors are assumed insignificant. While we cannot establish a functional relation between the distances, the figures do suggest that thresholds can be defined in the two domains such that the distance in coordinate space is normally under $d_{POSthres}$ (10 m, in our case) if the normalized distance in RSS space is below $d_{RSSthres}$ (8 dB, in our case). We use this information to form a constrained particle filter where positions with similar RSS (i.e. distance in RSS space below the threshold) can be used to constrain the current position estimation.

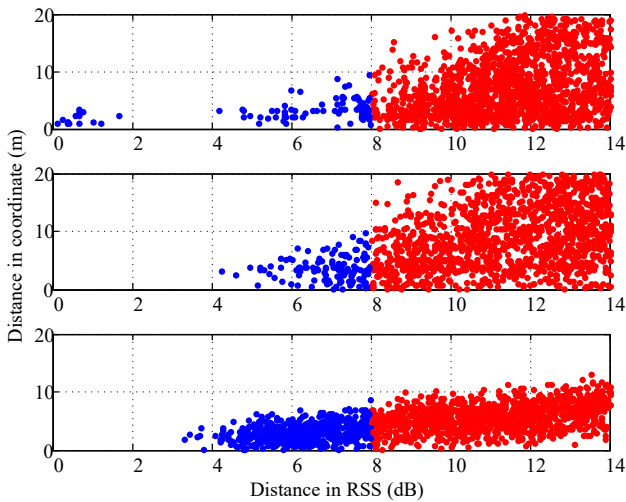

Fig. 7 Three samples sets of the distance in RSS space and in coordinate space. See text for discussion.

The weight update process based on similar RSS observations has three steps:

*1) Finding similar RSS.* Iterate over the observation list to find the RSS vectors whose normalized distance from the current RSS vector is less than $d_{RSSthres}$. Normally, we do not want to include the observations that are near in time or have been collected without walking in between, because they are not helpful for the loop-closure process. In our implementation, we only compare RSS values whose time difference is larger than 10 seconds and where the accumulated walking distance between them is larger than 20 meters. Again, these numbers are first choices which worked well for the examples herein but will be further investigated in the future.

We assume that there are $m$ RSS in the list satisfying the condition, namely $RSS_{n1}, RSS_{n2}, ..., RSS_{nm}$. The distances in RSS space are $d_{RSS,n1}, d_{RSS2,n2}, ...d_{RSS,nm}$ and the corresponding positions for the $i^{th}$ particle are $pos^i_{n1}, pos^i_{n2}, ..., pos^i_{nm}$.

*2) Estimating the current position by fingerprinting.* Use the weighted kNN algorithm to calculate a fingerprinting estimate of the current position as:

$$pos^i_{kNN} = \sum_{k=1}^{m} \frac{1}{d_{RSS,nk}} \frac{1}{NF} pos^i_{nk} \qquad (12)$$

where $NF$ is the normalized factor:

$$NF = \sum_{k=1}^{m} \frac{1}{d_{RSS,nk}} \qquad (13)$$

*3) Update weight.* If the distance between the estimated position $pos^i_{kNN}$ and the particle's current position $pos^i_k$ is larger than the threshold $d_{POSthres}$ then this particle is considered "impossible". However, to be more robust, the weight is set to be one percent of the original weight. Otherwise, the weigth of the particles is kept the same as in the previous epoch.

A simple example is given here to show the weight update process for the $i^{th}$ particle in our approach (Fig. 8). The two orange dots denote the positions which are similar in signal space found in step 1. The red dot denotes the estimated position through the weighted kNN algorithm at step 2. The blue dot denotes the current end position of the particle. $d^i_{kNN,k}$ is the distance between the two positions in coordinate space.

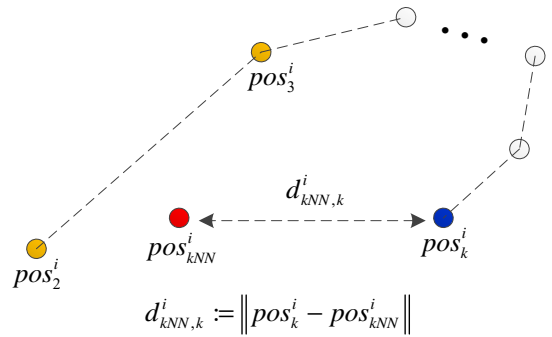

Fig. 8 A simple example of weight update process for the $i^{th}$ particle.

## IV. EXPERIMENT

### A. Experimental Setup

We have carried out real measurements to demonstrate the algorithm. Fig. 9 shows a scene during data acquisition. The inertial sensor is mounted on the foot of the user. The sensor is a multiple inertial measurement units (MIMU) platform with an embedded single-chip microcomputer [20]. The collected inertial data are processed in real-time in the foot module and the calculated positions are transmitted to the hand held Nexus 6P smart phone through Bluetooth Low Energy (BLE). The position estimation from the foot module is available step-wisely.

To obtain reference coordinates, a local coordinate frame was established using 15 prisms permanently mounted on the ceiling of the rooms. A total station Leica MS50 was utilized to measure the coordinates within this frame.

The hand held target is a frame (3-D printed) supporting the smart phone and a 360° mini-prism tracked by the total station. During the experiment the position of the cell phone is measured and tracked with the total station by the prism with high accuracy (a few mm) and update rate (about 1/0.07 s). The tracking process is controlled by a laptop running MATLAB. The position data collected by the total station are considered as ground truth within this experiment. A custom made app on the smartphone was used to collect RSS together with SSID and MAC of the available access points. These measurements were collected and stored by the application together with the readings from the foot-mounted module. The datasets collected by the Nexus 6P and the laptop were afterwards merged and synchronized manually in post-processing.

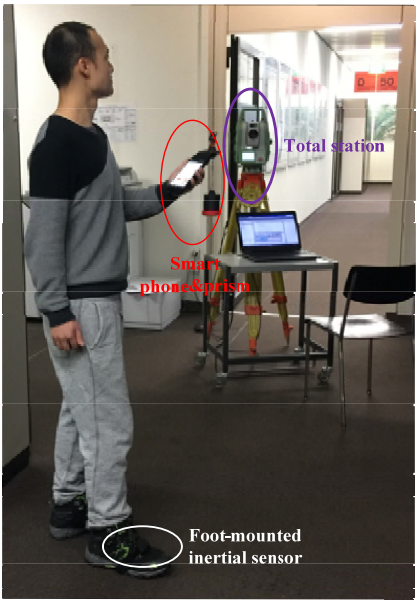

Fig. 9 The experimental scene.

### B. Results

As shown in Fig. 10, the total station (blue triangle) is placed at the corner of two perpendicular corridors, which can cover the areas in the black ellipse. The blue dot denotes the starting position of the user. As the black arrows show, the walking trajectories include walking outside and inside of the building. The user has walked the same trajectory for three iterations.

From the trajectories in Fig. 11, it is apparent that the new approach and the GP based approach have smaller positioning errors than the raw trajectory directly derived from the foot module. This shows the benefit of supporting the FMIP by signal strength measurements. The total walking length is about 1855 meters and the total time duration is about 24 minutes.

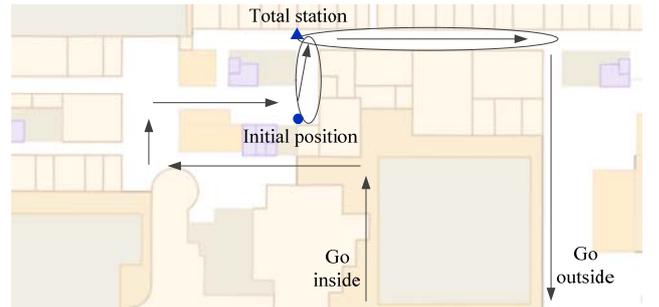

Fig. 10 The settings inside the building. The total station (blue triangle) is placed at the corner of two perpendicular corridors, which can cover the areas in the black ellipses. The blue dot denotes the starting position of the user. As the black arrows show, the walking trajectories include walking outside and inside of the building.

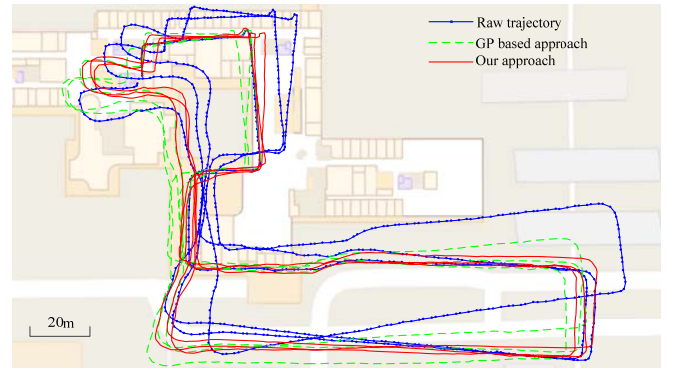

Fig. 11 Comparisons of the raw trajectory, GP based approach and our approach.

The user has re-visited the total station covered area twice. Therefore ground truth provided by the total station is available for the corresponding two parts of the walks and positioning errors can be calculated for these parts. Fig. 12 shows the errors for the first and second visits respectively. We can see that both the GP based approach and our approach improve the positioning accuracy. However, at the second re-visit, the error of the GP based approach becomes larger. The reasons may be two folds: wrong estimation of mean value of GP (positions of each AP is not known); and inconsistent weight update as mentioned before. Table I reports the mean errors from the two re-visits and further corroborates the results shown in Fig. 12.

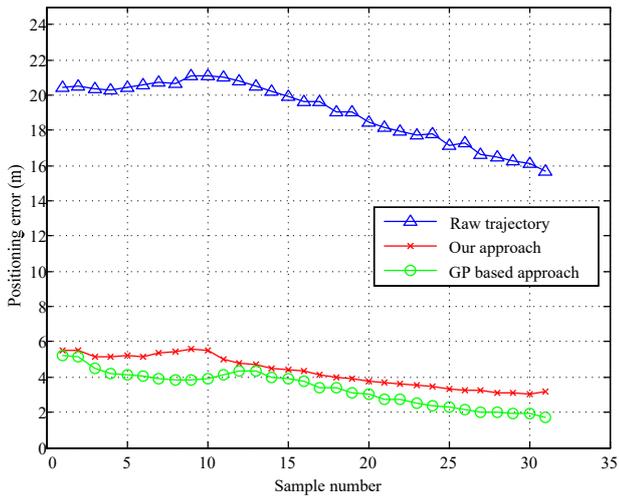

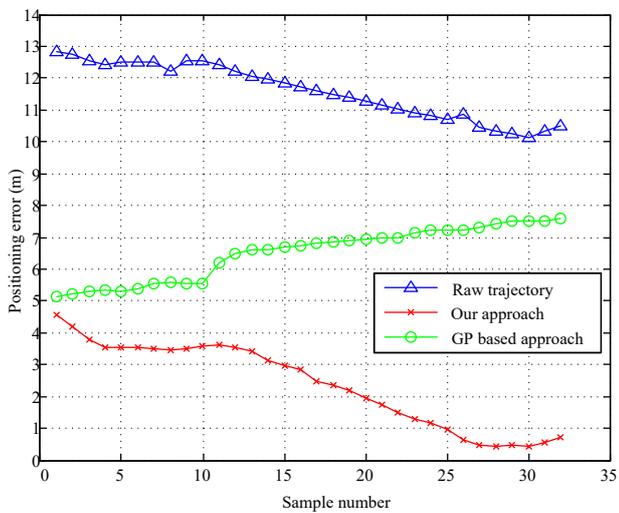

Fig. 12 Positioning error curves at the first re-visit (a) and the second re-visit (b) of the area covered by the total station.

Table II shows the processing time for the two approaches comprising the two visits covered by the total station. The two approaches both run on the same computer using MATLAB. The GP based approach consumes much more time than our approach. That is because in the GP based approach, at each filtering epoch, many independent GPs are trained independently and the number of samples in each GP grows larger with time. Noting that in each GP the computational complexity is $O(N^3)$, where $N$ is the number of samples used for training, it is clear that this approach cannot cope well with kinematic processing and large training sets.

TABLE I. MEAN ERRORS OF TWO RE-VISITS

| Mean Error | Trajectory | | |
|---|---|---|---|
| | *Raw trajectory* | *GP based approach* | *Our approach* |
| First re-visit | 19.1m | 3.4m | 4.3m |
| Second re-visit | 11.6m | 6.5m | 2.4m |

TABLE II. PROCESSING TIME OF THE TWO APPROACHES

| **Processing time** | *GP based approach* | *Our approach* |
|---|---|---|
| First re-visit | 1179s | 29s |
| Second re-visit | 6887s | 96s |

## V. CONCLUSIONS

A particle filter based indoor positioning approach combining FMIP and FWIP is proposed in this paper. This approach works without any labor-intensive site survey phase, and can greatly suppress positioning error growth (due to inertial drifts of the foot-mounted IMU) with time. Compared with the GP based approaches, the experiment shows our approach can have a better performance in terms of both positioning accuracy and processing time.

Our future work will continue in two ways: (1) the impact of different choices of parameters (e.g., thresholds in determine "similar" RSS) will be investigated; (2) the approach will be extended for not only estimating the positions but also the corresponding radio map.

ACKNOWLEDGMENT

The Chinese Scholarship Council has supported Y.G. during his academic visit at ETH and C.Z. during his Ph.D. studies at ETH.